\documentclass[pra,reprint]{revtex4-1}
\usepackage{graphicx}
\usepackage{appendix}
\usepackage{braket}
\usepackage{rotating}
\usepackage{amssymb}
\usepackage{amsmath}
\usepackage{txfonts}
\usepackage{bm}
\usepackage{color}
\usepackage{multirow}
\usepackage{booktabs}
\usepackage{dcolumn}
\usepackage{subcaption}

\begin{document}

%\title{Probability-theoretical estimates of the diameters of the Rubik's Cube groups}
\title{Probabilistic estimates of the diameters of the Rubik's Cube groups}

\author{So \surname{Hirata}}
\email{sohirata@illinois.edu}
\affiliation{Department of Chemistry, University of Illinois at Urbana-Champaign, Urbana, Illinois 61801, USA}

\date{\today}

\begin{abstract}
The diameter of the Cayley graph of the Rubik's Cube group is the fewest number of turns needed to solve the Cube from the hardest initial configuration.
For the 2$\times$2$\times$2 Cube, the diameter is 11 in the half-turn metric, 14 in the quarter-turn metric, 19 in the semi-quarter-turn metric,
and 10 in the bi-quarter-turn metric. 
For the 3$\times$3$\times$3 Cube, the diameter was determined by Rokicki {\it et al.}\  to be 20 in the half-turn metric and 26 in the quarter-turn metric. 
This study shows that a modified version of the coupon collector's problem in probability theory can predict the diameters correctly for both  2$\times$2$\times$2 
and 3$\times$3$\times$3 Cubes insofar as the quarter-turn metric is adopted. In the half-turn metric, the diameters are overestimated by one and two, respectively, for the 
2$\times$2$\times$2 
and 3$\times$3$\times$3 Cubes, whereas for the 2$\times$2$\times$2 Cube
in the semi-quarter-turn and bi-quarter-turn metrics, they are overestimated by two and underestimated by one, respectively. 
Invoking the same probabilistic logic, the diameters of the 4$\times$4$\times$4 and 5$\times$5$\times$5 Cubes are predicted to be 48 (41) and 68 (58) in the quarter-turn (half-turn) metric, whose precise determinations are far beyond reach of classical supercomputing. 
The probabilistically estimated diameter is shown to obey the approximate formula of $\ln N / \ln r + \ln N / r$, where $N$ is the number of configurations and $r$ is the branching ratio. %The same formula approximates the diameter of a cubic symmetric graph with $N$ vertexes and girth $g$, where $r = k - k^{2-g}$. 
\end{abstract}

\maketitle 

%==============================
\section{Objective}
%==============================

The diameter of the Cayley graph of the Rubik's Cube group \cite{Hofstadter,Singmasterbook,Joyner2008,rewix} (hereafter the diameter) is 
the fewest number of turns %(also called twists or rotations or moves) 
to solve the Cube from the hardest initial configuration. 
This number is also colloquially referred to as God's Number.
Here, a configuration denotes a unique arrangement of sticker colors of the Cube reachable 
 by a series of turns from the solved configuration, while holding the orientation of the whole Cube 
 %(or of  the six central facelets in the case of the 3$\times$3$\times$3 Rubik's Cube) 
 fixed.

After decades of research by the community of mathematicians 
and computer scientists \cite{thistle1981,kochiemba2024,Kunkle2007,Kunkle2009}, culminating in a Herculean computational effort by Rokicki {\it et al.}\ \cite{Rokicki2013,Rokicki2014,cube20}, the diameter of  the 3$\times$3$\times$3 Cube 
was determined to be 20 in the half-turn metric or 26 in the quarter-turn metric \cite{cube20,Rokicki2013,Rokicki2014}.
The corresponding number for the 2$\times$2$\times$2 Cube is 11 in the half-turn metric and 14 in the quarter-turn metric \cite{rewix}.
%Here, a ``metric'' means the definition of an allowed turn, such as a clockwise 90$^\circ$  or counterclockwise rotation of each of the right, left, front, back, top or down layer in the case of the quarter-turn metric. 
See below for the definitions of different metrics and other technical details.

The objective of this study is to examine to what extent
a simple probabilistic argument can predict the diameters for different Rubik's Cube groups in various metrics,
and to offer our best probabilistic estimates of the diameters for the 4$\times$4$\times$4 and 5$\times$5$\times$5 Cubes, whose
brute-force determination is presently unthinkable.

%==============================
\section{Background}
%==============================

Rubik's Cube \cite{Hofstadter,Singmasterbook,Joyner2008,rewix} has long been a popular testbed for computer science research since the pioneering 
study by Korf \cite{Korf1983thesis,Korf1985}, 
especially in the areas of artificial intelligence and optimization theory. It has also fascinated mathematicians working on group theory \cite{Zassenhaus1982,Fiat1989,Joyner2008,Demaine2011,Ryabogin2012,Bonzio2017,Bonzio2018} and graph theory \cite{Beenker1988,Bollobasbook,Biggsbook,Ferrero2014,Westbook}.
Furthermore, its similarities with problems appearing in quantum mechanics \cite{Corli2021}, statistical mechanics \cite{Lee2008,Chen2012,Chen2014},
high-energy physics \cite{Czech2011}, %microfluidic device,\cite{Lai2020} 
and cryptography \cite{Diaconu2013,Ionescu2015} have been noted. 
 
The research on Rubik's Cube may be largely divided into three groups by their distinct objectives. 
The first group aims at determining the diameters described above as a mathematical question. %, which is a mathematical question.
The present study belongs to this group.

The second group's objective is to discover or develop an optimization method that is powerful enough to solve the Cube in any configurations 
 efficiently. The method may also be general enough to do so without any domain knowledge, so that it can 
be applied to other puzzles. Machine Learning algorithms tend to fail \cite{Lichodzijewski2011,Smith2016,Johnson2018} because 
a required fitness function is  hard to identify or define. %, namely, for the same reason why Rubik's Cube is a hard puzzle for the most intelligent human player.
Successful strategies first discover a good fitness function computationally \cite{Johnson2021} and/or to
rely on large pattern databases \cite{Korf97,Culberson1998,Korf2002,Arfaee2011},
both of which are not meant to help a human player. 
So far the most effective method---the reinforcement learning algorithm recently developed by Agostinelli {\it et al.}\ \cite{Agostinelli2019}---can 
solve the 3$\times$3$\times$3 Cube 100\% of cases; with 60\% of cases it discovers the most efficient solutions, underscoring its potential utility in determining 
the diameters of higher-order Cubes. 

The third group computationally searches for the so-called ``macros'' or a series of turns that a human player can potentially master in a stepwise solution of the Cube. 
Such a solution is neither expected nor required to be efficient, but it may possibly contain some general principles or strategies 
for solutions, which may be applied to higher-order Cubes or other related puzzles. %If such principles or strategies can  indeed be extracted either manually or computationally, they may be argued to be a more useful form of computer intelligence for human players. 
These studies, therefore, belong to the developments of expert systems \cite{Jackson1998} or discovery systems \cite{SHEN1990257,Gil2014}.
Korf's groundbreaking study \cite{Korf1985,Korf1983thesis} falls into this category, which exhaustively and systematically lists macros
for both 2$\times$2$\times$2 and 3$\times$3$\times$3 Cubes. As a by-product, they also set the (albeit conservative) upper bounds for the diameters, spurring many subsequent 
studies. 

%==============================
\section{Probabilistic estimate}
%==============================

Let $N$ be the total number of configurations. 
Consider a brute-force enumeration of configurations by performing all possible turns at each step, starting from the initial, solved configuration.
Generated configurations are recorded in a non-duplicate list, and after a certain number of steps, all $N$ configurations have been generated and no new configuration
can be added to the list. This number is equal to the diameter. 
This brute-force algorithm is as faultless as it is intractable for all but the 2$\times$2$\times$2 Cube. 

We shall instead estimate this number by simulating the above brute-force algorithm under the assumption
that all generated configurations are random and independent of the previous ``seed'' configurations (this non-Markovian assumption 
is clearly not obeyed in reality and is a major approximation). 
We can then view this problem as a modified version of the 
coupon collector's problem \cite{Gumbel1941,Dawkins1991,Flajolet1992,Blom1994} in probability theory (see Appendix \ref{sec:coupon}).
See pages 26 and 34 of Singmaster's book \cite{Singmasterbook}, where the author develops a similar argument.
It is also related to mathematical models of epidemics \cite{Hill2020} and network theory \cite{Newman2008}. 

The coupon collector's problem asks the following question:\ Let $T_N$ be the number of times an integer in the range of 1 through $N$ is generated uniformly randomly
 until every integer 1 through $N$ is realized at least once. What is the expectation value of $T_N$? The answer is 
\begin{eqnarray}
E[T_N] = N \ln N + \gamma N, \label{coupon}
\end{eqnarray}
where $\gamma=0.5772$ is the Euler--Mascheroni constant. 
The standard deviation of $T_N$ is
\begin{eqnarray}
\sigma[T_N] = 1.28\, N. \label{sigma}
\end{eqnarray}
See  Appendix \ref{sec:coupon} for derivations. 

The central limit theorem does not apply to $T_N$, whose distribution is said to be the Siobhan distribution \cite{Dawkins1991}, which is related to the Gumbel distributions \cite{Gumbel1941}.
As per Dawkins \cite{Dawkins1991}, the probability that all integers are generated after random generations have been repeated $T$ times is given asymptotically by
\begin{eqnarray}
\text{Pr}\left(T_N \leq T\right) \approx e^{ -N e^{-T/N}}. \label{siobhan}
\end{eqnarray}

The difference between the coupon collector's problem and the probabilistic estimation of the diameters is that in the latter 
the random configurations are generated in batches or stepwise. We therefore propose modifying the original problem as follows:

Let $C(t)$ be the number of configurations generated in the $t$th step by performing all possible $k$ turns to each of  non-redundant ``seed'' configurations
in the previous step $t-1$. Let the number of these seed configurations be $S(t-1)$.
Here, ``non-redundant'' means that no two configurations in the set are the same (lest the configurations generated from them are highly redundant 
and less uniformly random). 

In the $t$th step, from each of the $S(t-1)$ non-redundant seed configurations,
$r$ new configurations are generated by performing all $k$ turns, but excluding some combinations
of turns (``spurious turns''; see the following sections for examples) that immediately give rise to duplicate configurations:
\begin{eqnarray}
C(t) = r\, S(t-1). \label{growth}
\end{eqnarray}
Here, $r$ (the branching ratio) 
is a real positive number that is less than $k$, and this condition $r < k$ effectively excludes the spurious turns.  The value of $r$ is 
determined from the average growth ratio of the number of configurations generated in the first few steps (see the following sections). 
The purpose of this exclusion is not to eliminate any duplicate entries in the set of $C(t)$ newly generated configurations (the set is supposed to be 
redundant and increaingly more so as its size approaches or exceeds $N$); rather, it is to minimize the hysteresis in the generation and to make the generated configurations
more uniformly random. 

Collecting only the distinct configurations among the $C(t)$ generated ones, we obtain a new set of seed configurations
used in the next step. The size $S(t)$ of this non-redundant set is related to the size $C(t)$ of the original set by 
\begin{eqnarray}
S(t) &=& N \left( 1 - e^{-C(t)/N} \right),\label{deltaSlargen} 
\end{eqnarray}
whose derivation is given in Appendix \ref{sec:derivation2}. The seed configurations should be non-redundant in 
that no two configurations in it are the same, ensuring more uniform randomness of the generated configurations, but 
some of them are certainly repeated from the seed sets of previous steps.

Because the seed set is non-redundant, its size $S(t)$ cannot exceed $N$. This is ensured by Eq.\ (\ref{deltaSlargen}). 
Furthermore, in earlier steps, $C(t) \ll N$ and because the use of $r$ effectively eliminates spurious turns, there is a vanishing probability that 
two randomly generated configurations happen to be the same, implying $S(t) \approx C(t)$. This is also consistent with the first-order Taylor-series approximation
of the exponential in Eq.\ (\ref{deltaSlargen}).

The total number of redundantly generated configurations,
\begin{eqnarray}
T(t) &=&  C(0) + \dots +  C(t), \label{G}
\end{eqnarray} 
will be compared with $E[T_N]$. When $T(t)$ exceeds $E[T_N]$ for the first time, $t$ is the probabilistic best estimate of the diameter. 
Nowhere in this estimation do actual configurations need to be generated; nor is a random number needed. The whole analysis can in principle proceed by computing only the 
estimated sizes of various sets of configurations. 

%==============================
\section{The 2$\times$2$\times$2 Cube}
%==============================

One of the eight corner cubies (say, the front-up-left one) of the 2$\times$2$\times$2 Cube is held fixed as the orientation anchor of the whole Cube
(here, a ``cubie'' stands for a 1$\times$1$\times$1 constituent cube). 
The remaining seven cubies' positions can be permuted in $7!$ ways and their orientations can take $3^7$ ways, except the last (seventh) cubie's orientation
is fixed by the first six (it is said \cite{Hofstadter} that the ``sanity'' of the seven cubies is maintained by only one of three orbits \cite{Joyner2008}). 
Therefore, the total number of configurations $N$ (or the order of the 2$\times$2$\times$2 Rubik's Cube group) is
\begin{eqnarray}
N = \frac{7! \cdot 3^7}{3} = 3.67 \times 10^{6}. \label{N2}
\end{eqnarray}

According to Eqs.\ (\ref{coupon}) and (\ref{sigma}), the expected number of random configurations  $E[T_N]$
that need to be generated before each of $N$ configurations occurs at least once (and its standard deviation $\sigma[T_N]$) is
\begin{eqnarray}
E[T_N] \pm \sigma[T_N] =  \Big( 15.7\pm1.3\Big) \,N.
\end{eqnarray}
 It thus suggests that before the most stubborn one of all 3.67 million configurations is finally 
realized, an average configuration has to recur as many as 16 times on average.

\subsection{Half-turn metric}

With the front-up-left cubie held fixed, only the right ($R$), downward ($D$) and back ($B$) layers can be rotated. 
In the half-turn metric, the Cube can undergo one of the three (3) clockwise 90$^\circ$ turns ($R$, $D$, $B$), three (3) counterclockwise 90$^\circ$ turns ($R^{-1}$, $D^{-1}$, $B^{-1}$), and three (3) 180$^\circ$ turns ($R^2$, $D^2$, $B^2$) in the Singmaster notation \cite{Singmasterbook}.

In the first step, all nine (9) turns lead to distinct configurations. 
In the second and subsequent steps, $X$, $X^{-1}$, and $X^2$ immediately following the same (where $X=R$, $D$, or $B$)
only produce the configurations that have already occurred. Therefore, $r \leq 6$. An explicit enumeration indicates 
\begin{eqnarray}
S(0) = C(0) &=& 1, \\
S(1) = C(1) &=& 9, \\
S(2) = C(2) &=& 54,\\
S(3) = C(3) &=& 321.
\end{eqnarray}
The precise evaluation of $S(t)$ at arbitrary value of $t$ is still feasible for the 2$\times$2$\times$2 Cube, 
and it is tantamount to determining the diameter by a brute-force enumeration.

Instead, starting with $t=4$, we approximate the number of redundant random configurations $C(t)$ in the $t$th step generated 
by performing all possible turns to the $S(t-1)$ non-redundant seed configurations in the previous step, 
excluding the turns that predictably and immediately lead to the already realized configurations. 
Adopting the branching ratio of $r = 321/54 = 5.94$ to effectuate this exclusion, we use Eq.\ (\ref{growth}) to determine $C(t)$ from $S(t-1)$. %, which is reproduced.
%\begin{eqnarray}
%C(t) &=& r\,S(t-1).  \nonumber
%\end{eqnarray}
The seed configurations of the subsequent ($t$th) step are then obtained by deleting all duplicate entries in the $C(t)$ set.
The size of this set $S(t)$ is estimated by Eq.\ (\ref{deltaSlargen}). %, which is also reproduced.
%\begin{eqnarray}
% S(t) &=& N \left( 1 - e^{-C(t)/N} \right), \nonumber
%\end{eqnarray}
%This set of seed configurations is non-redundant in the sense 
%that no two configurations in it are the same, although some of them are repeated from the seed configuration sets of previous steps.
%Its size, therefore, cannot exceed $N$, which is ensured by Eq.\ (\ref{deltaSlargen}). 
%In earlier steps, where $C(t) \ll N$, the first-order Taylor expansion suggests $S(t) \approx C(t)$, which is consistent with Eqs.\ (\ref{s0})--(\ref{s3}).
The total number of redundantly generated configurations $T(t)$ is then given by Eq.\ (\ref{G}) and will be compared with $E[T_N]$.
%\begin{eqnarray}
%T(n) &=&  C(0) + \dots +  C(t),
%\end{eqnarray} 
%will be compared with $M$. 

\begin{table}
\caption{\label{table:2x2x2half} The probabilistic estimates of the numbers of non-redundant seed configurations $S(t)$,   redundant configurations $C(t)$,
and cumulative  redundant configurations $T(t)$ %, cumulative non-redundant configurations $T(t)$, and newly generated non-redundant configurations $N(n)$
as a function of step ($t$) in the half-turn metric of the 2$\times$2$\times$2 Cube. 
$N = 3.67\times10^6$, $E[T_N]  = 15.7\,N$, and therefore the predicted diameter is 12 for $T(12) > E[T_N] > T(11)$, while the correct diameter is 11.}
\begin{ruledtabular}
\begin{tabular}{rrrr}
$t$ & $S(t)$  & $C(t)$ & $T(t)$ \\ \hline %& $T(t)$ & $N(n)$ \\ \hline  
0 & 	1 & 	1 & 	1 \\ % & 	1 & 	1 \\
1 & 	9 & 	9 & 	10 \\ % & 	10 & 	9 \\
2 & 	54 & 	54 & 	64 \\ % & 	64 & 	54 \\
3 & 	321 & 	321 & 	385 \\ % & 	385 & 	321 \\
4 & 	0.001\,$N$ & 	0.001\,$N$ & 	0.001\,$N$ \\ % & 	2291 & 	1906 \\
5 & 	0.003\,$N$ & 	0.003\,$N$ & 	0.004\,$N$ \\ % & 	13590 & 	11299 \\
6 & 	0.018\,$N$ & 	0.018\,$N$ & 	0.022\,$N$ \\ % & 	79889 & 	66300 \\
7 & 	0.102\,$N$ & 	0.108\,$N$ & 	0.130\,$N$ \\ % & 	446501 & 	366612 \\
8 & 	0.454\,$N$ & 	0.606\,$N$ & 	0.735\,$N$ \\ % & 	1913158 & 	1466657 \\
9 & 	0.933\,$N$ & 	2.699\,$N$ & 	3.435\,$N$ \\ % & 	3555711 & 	1642553 \\
10 & 	0.996\,$N$ & 	5.540\,$N$ & 	8.975\,$N$ \\ % & 	3673695 & 	117985 \\
11 & 	0.997\,$N$ & 	5.917\,$N$ & 	14.892\,$N$ \\ % & 	3674159 & 	464 \\
12 & 	0.997\,$N$ & 	5.924\,$N$ & 	20.816\,$N$ \\ % & 	3674160 & 	1 \\
%13 & 	0.997\,$N$ & 	5.924\,$N$ & 	26.740\,$N$ \\ % & 	3674160 & 	0 \\
\end{tabular}
\end{ruledtabular} 
\end{table}

Table \ref{table:2x2x2half} lists $S(t)$, $C(t)$, and $T(t)$ thus computed 
as a function of $t$ up to $t=12$, where the cumulative number of configurations $T(12) = 20.8\,N$
exceeds $E[T_N] = 15.7\,N$ for the first time. Therefore, the predicted diameter of the 2$\times$2$\times$2 Cube in the half-turn metric is 12, which overshoots the 
correct diameter of 11 \cite{rewix}.

According to the Siobhan distribution [Eq.\ (\ref{siobhan})], the probabilities that all configurations have been realized at $t=11$ and 12 are
\begin{eqnarray}
\text{Pr}(T_N \leq T(11))&=& 0.286, \\
\text{Pr}(T_N \leq T(12)) &=& 0.997,
\end{eqnarray}
respectively. Hence, the present probabilistic argument predicts that there is a fair (28.6\%) chance that the diameter is correctly 11,
which is consistent with the fact that $T(11)$ is within one standard deviation $\sigma[T_N]$ from $E[T_N]$; the predicted diameter's  overestimation is not severe. 
Correspondingly, according to Appendix \ref{sec:derivation2}, the expected numbers of configurations that have {\it not} been realized after $t=11$ and 12 are
\begin{eqnarray}
N e^{-T(11)/N} &=& 1.3, \\
N e^{-T(12)/N} &=& 0.003,
\end{eqnarray}
respectively. 

%\begin{figure} \\
%	\includegraphics[width=\columnwidth]{Figures/half.png}
%	\caption{The  actual and predicted numbers of new non-redundant configurations generated as a function of step in the half-turn metric.
%	The terminal plot is the last step at which nonzero new configurations are generated and corresponds to the actual or 
%	most conservatively predicted diameter.\label{fig:half}}
%\end{figure} 

\subsection{Quarter-turn metric\label{sec:2x2x2Q}}

In this metric, there are only six (6) valid 90$^\circ$ turns: $R$, $D$, $B$ and their inverses $R^{-1}$, $D^{-1}$, $B^{-1}$. 
An explicit evaluation gives
\begin{eqnarray}
S(0) = C(0) &=& 1, \\
S(1) = C(1) &=& 6, \\
S(2) = C(2) &=& 27,\\
S(3) = C(3) &=& 120,
\end{eqnarray}
suggesting $r = 120/27 = 4.44$. Using this $r$ in conjunction with Eqs.\ (\ref{growth})--(\ref{G}), we obtain
Table \ref{table:2x2x2quarter}. $T(t)$ exceeds $E[T_N] = 15.7\,N$ for the first time at $t=14$, and therefore the predicted diameter
of the 2$\times$2$\times$2 Cube in the quarter-turn metric is 14, which agrees with the correct diameter of 14 \cite{rewix}.

\begin{table}
\caption{\label{table:2x2x2quarter} Same as Table \ref{table:2x2x2half} but in the quarter-turn metric.
The predicted diameter is 14 for $T(14) > E[T_N]  > T(13)$, while the correct diameter is also 14.}
\begin{ruledtabular}
\begin{tabular}{rrrr}
$t$ & $S(t)$  & $C(t)$ & $T(t)$ \\ \hline %& $T(t)$ & $N(n)$ \\ \hline
0 & 	1 & 	1 & 	1 \\ % & 	1 & 	1 \\
1 & 	6 & 	6 & 	7 \\ %& 	7 & 	6 \\
2 & 	27 & 	27 & 	34 \\ %& 	34 & 	27 \\
3 & 	120 & 	120 & 	154 \\ %& 	154 & 	120 \\
4 & 	533 & 	533 & 	687 \\ %& 	687 & 	533 \\
5 & 	0.001\,$N$ & 	0.001\,$N$ & 	0.001\,$N$ \\ %& 	3051 & 	2364 \\
6 & 	0.003\,$N$ & 	0.003\,$N$ & 	0.004\,$N$ \\ %& 	13531 & 	10480 \\
7 & 	0.013\,$N$ & 	0.013\,$N$ & 	0.016\,$N$ \\ %& 	59701 & 	46170 \\
8 & 	0.055\,$N$ & 	0.056\,$N$ & 	0.073\,$N$ \\ %& 	257763 & 	198062 \\
9 & 	0.221\,$N$ & 	0.250\,$N$ & 	0.323\,$N$ \\ %& 	1014051 & 	756288 \\
10 & 	0.626\,$N$ & 	0.983\,$N$ & 	1.306\,$N$ \\ %& 	2678666 & 	1664615 \\
11 & 	0.938\,$N$ & 	2.778\,$N$ & 	4.084\,$N$ \\ %& 	3612303 & 	933637 \\
12 & 	0.984\,$N$ & 	4.164\,$N$ & 	8.248\,$N$ \\ %& 	3673199 & 	60895 \\
13 & 	0.987\,$N$ & 	4.371\,$N$ & 	12.619\,$N$ \\ %& 	3674148 & 	949 \\
14 & 	0.988\,$N$ & 	4.384\,$N$ & 	17.003\,$N$ \\ %& 	3674160 & 	12 \\
15 & 	0.988\,$N$ & 	4.385\,$N$ & 	21.388\,$N$ \\ %& 	3674160 & 	12 \\
\end{tabular}
\end{ruledtabular} 
\end{table}

As per Eq.\ (\ref{siobhan}), the probabilities that all configurations have been generated by $t= 13$, 14, and 15 are, respectively, 
\begin{eqnarray}
\text{Pr}(T_N \leq T(13))&=& 5 \times 10^{-6}, \\
\text{Pr}(T_N \leq T(14)) &=& 0.859, \\
\text{Pr}(T_N \leq T(15)) &=& 0.998. 
\end{eqnarray}
Correspondingly, the estimated numbers of the unrealized configurations at $t=13$, 14, and 15 are
\begin{eqnarray}
N e^{-T(13)/N} &=& 12, \\
N e^{-T(14)/N} &=& 0.15, \\
N e^{-T(15)/N} &=& 0.002.
\end{eqnarray}
Hence, there is a negligible chance that the predicted diameter is actually 13 (not 14), but a small (14\%) probability that it is 15. In this case, the predicted
most likely (86\%) value of the diameter (14) proves to be exact. 
%\begin{figure} \\
%	\includegraphics[width=\columnwidth]{Figures/quarter.png}
%	\caption{The same as Fig.\ \ref{fig:half} but in the quarter-turn metric.}
%\end{figure} 

\subsection{Semi-quarter-turn metric}

In this newly introduced metric, only three (3) clockwise 90$^\circ$ turns ($R$,  $D$, and $B$) are allowed in each step.
Since the inverses are not included, its ``group'' may not meet the mathematical definition of a group and the corresponding graph may not be a Cayley graph.
Nevertheless, all of $N$ configurations can be reached by combinations of these three turns and the ``diameter'' is still well defined.

With an explicit enumeration, we find
\begin{eqnarray}
S(0) = C(0) &=& 1, \\
S(1) = C(1) &=& 3, \\
S(2) = C(2) &=& 9,\\
S(3) = C(3) &=& 27, \\
S(4) = C(4) &=& 78, \\
S(5) = C(5) &=& 216, 
\end{eqnarray}
suggesting $r = 216/78 = 2.77$. Using Eqs.\ (\ref{growth})--(\ref{G}) with this $r$ for $t \geq 6$, we obtain Table \ref{table:2x2x2semiquarter}. 
$T(21)$ exceeds $E[T_N]$ for the first time and therefore the predicted diameter is 21, which is too large as compared with
the correct diameter of 19. 

\begin{table}
\caption{\label{table:2x2x2semiquarter} Same as Table \ref{table:2x2x2half} but in the semi-quarter-turn metric.
The predicted diameter is 21 for $T(21) > E[T_N] > T(20)$, while the correct diameter is 19.}
\begin{ruledtabular}
\begin{tabular}{rrrr}
$t$ & $S(t)$  & $C(t)$ & $T(t)$ \\ \hline % & $T(t)$ & $N(n)$ \\ \hline 
0 & 	1 & 	1 & 	1 \\ %& 	1 & 	1 \\
1 & 	3 & 	3 & 	4 \\ %& 	4 & 	3 \\
2 & 	9 & 	9 & 	13 \\ %& 	13 & 	9 \\
3 & 	27 & 	27 & 	40 \\ %& 	40 & 	27 \\
4 & 	78 & 	78 & 	118 \\ %& 	118 & 	78 \\
5 & 	216 & 	216 & 	334 \\ %& 	334 & 	216 \\
6 & 	583 & 	583 & 	917 \\ %& 	932 & 	598 \\
7 & 	1546 & 	1546 & 	2463 \\ %& 	2589 & 	1656 \\
8 & 	0.001\,$N$ & 	0.001\,$N$ & 	0.002\,$N$ \\ %& 	7172 & 	4583 \\
9 & 	0.003\,$N$ & 	0.003\,$N$ & 	0.005\,$N$ \\ %& 	19830 & 	12658 \\
10 & 	0.010\,$N$ & 	0.010\,$N$ & 	0.015\,$N$ \\ %& 	54606 & 	34775 \\
11 & 	0.026\,$N$ & 	0.026\,$N$ & 	0.041\,$N$ \\ %& 	148770 & 	94165 \\
12 & 	0.070\,$N$ & 	0.072\,$N$ & 	0.113\,$N$ \\ %& 	393883 & 	245113 \\
13 & 	0.175\,$N$ & 	0.193\,$N$ & 	0.306\,$N$ \\ %& 	968527 & 	574644 \\
14 & 	0.384\,$N$ & 	0.485\,$N$ & 	0.791\,$N$ \\ %& 	2008731 & 	1040204 \\
15 & 	0.655\,$N$ & 	1.065\,$N$ & 	1.856\,$N$ \\ %& 	3100011 & 	1091281 \\
16 & 	0.837\,$N$ & 	1.815\,$N$ & 	3.671\,$N$ \\ %& 	3580672 & 	480661 \\
17 & 	0.902\,$N$ & 	2.319\,$N$ & 	5.990\,$N$ \\ %& 	3664963 & 	84291 \\
18 & 	0.918\,$N$ & 	2.498\,$N$ & 	8.488\,$N$ \\ %& 	3673403 & 	8440 \\
19 & 	0.921\,$N$ & 	2.542\,$N$ & 	11.030\,$N$ \\ %& 	3674100 & 	697 \\
20 & 	0.922\,$N$ & 	2.552\,$N$ & 	13.582\,$N$ \\ %& 	3674155 & 	55 \\
21 & 	0.922\,$N$ & 	2.554\,$N$ & 	16.136\,$N$ \\ %& 	3674160 & 	4 \\
22 & 	0.922\,$N$ & 	2.555\,$N$ & 	18.691\,$N$ \\ %& 	3674160 & 	0 \\
\end{tabular}
\end{ruledtabular}
\end{table}

The probabilities that all configurations have been generated by $t= 20$, 21, and 22 are, respectively, 
\begin{eqnarray}
\text{Pr}(T_N \leq T(20))&=& 0.0097, \\
\text{Pr}(T_N \leq T(21)) &=& 0.697, \\
\text{Pr}(T_N \leq T(22)) &=& 0.972,
\end{eqnarray}
with the estimated numbers of the non-generated configurations being
\begin{eqnarray}
N e^{-T(20)/N} &=& 4.6, \\
N e^{-T(21)/N} &=& 0.36, \\
N e^{-T(22)/N} &=& 0.03.
\end{eqnarray}
In this case, the probabilistic estimate misses the mark completely, predicting that there is zero probability that the diameter is correctly
19. 

%\begin{figure} \\
%	\includegraphics[width=\columnwidth]{Figures/semiquarter.png}
%	\caption{The same as Fig.\ \ref{fig:half} but in the semi-quarter-turn metric.}
%\end{figure} 

\subsection{Bi-quarter-turn metric}

In this another new metric, all nine (9) turns in the half-turn metric plus six (6) additional bi-quarter turns ($RD$, $R^{-1} D^{-1}$, $DB$, $D^{-1} B^{-1}$, $BR$, $B^{-1} R^{-1}$)
are elementary operations. This metric is intentionally highly redundant. Lacking some inverses, its ``group'' and ``Cayley graph'' may not meet 
the proper mathematical definitions, but we still use the ``diameter'' to refer to the fewest number of turns to solve the Cube.

In the first few steps, the numbers of configurations are
\begin{eqnarray}
S(0) = C(0) &=& 1, \\
S(1) = C(1) &=& 15, \\
S(2) = C(2) &=& 144,\\
S(3) = C(3) &=& 1324.
\end{eqnarray}
Using $r = 1324/144 = 9.19$, we can generate the numbers of configurations  compiled in Table \ref{table:2x2x2biquarter}.
The first time $T(t)$ exceeds $E[T_N]$ is $t=9$, predicting the diameter of 9, which is too small as compared with the correct diameter of 10. 

\begin{table}
\caption{\label{table:2x2x2biquarter} Same as Table \ref{table:2x2x2half} but in the bi-quarter-turn metric.
The predicted diameter is 9 for $T(9) > E[T_N] > T(8)$, while the correct diameter is 10.}
\begin{ruledtabular}
\begin{tabular}{rrrr}
$t$ & $S(t)$  & $C(t)$ & $T(t)$ \\ \hline % & $T(t)$ & $N(n)$ \\ \hline 
0 & 	1 & 	1 & 	1 \\ %& 	1 & 	1 \\
1 & 	15 & 	15 & 	16 \\ %& 	16 & 	15 \\
2 & 	144 & 	144 & 	160 \\ %& 	160 & 	144 \\
3 & 	1324 & 	1324 & 	1484  \\ %& 	1484 & 	1324 \\
4 & 	0.003\,$N$ & 	0.003\,$N$ & 	0.004\,$N$ \\ %& 	13626 & 	12143 \\
5 & 	0.030\,$N$ & 	0.030\,$N$ & 	0.034\,$N$ \\ %& 	123174 & 	109548 \\
6 & 	0.240\,$N$ & 	0.275\,$N$ & 	0.309\,$N$ \\ %& 	977005 & 	853831 \\
7 & 	0.890\,$N$ & 	2.210\,$N$ & 	2.519\,$N$ \\ %& 	3378199 & 	2401194 \\
8 & 	1.000\,$N$ & 	8.182\,$N$ & 	10.700\,$N$ \\ %& 	3674077 & 	295878 \\
9 & 	1.000\,$N$ & 	9.187\,$N$ & 	19.888\,$N$ \\ %& 	3674160 & 	83 \\
10 & 	1.000\,$N$ & 	9.189\,$N$ & 	29.077\,$N$ \\ %& 	3674160 & 	0 \\
11 & 	1.000\,$N$ & 	9.189\,$N$ & 	38.266\,$N$ \\ %& 	3674160 & 	0 \\
\end{tabular}
\end{ruledtabular}
\end{table}

As per Eq.\ (\ref{siobhan}), 
\begin{eqnarray}
\text{Pr}(T_N \leq T(9))&=& 0.992, \\
\text{Pr}(T_N \leq T(10)) &=& 0.999999, 
\end{eqnarray}
implying that there is slightly less than 1\% chance that the diameter is 10.
No configuration has been unrealized by $t=9$ since
\begin{eqnarray}
N e^{-T(9)/N} &=& 0.008,
\end{eqnarray}
which is also inferred from the fact that $T(9)$ exceeds $E[T_N]$ by $3\sigma[T_N]$. 

%==============================
\section{The 3$\times$3$\times$3 Cube}
%==============================

The central cubie on each of the six faces of the 3$\times$3$\times$3 Cube is held fixed as the orientation anchor. The number of configurations is \cite{Hofstadter}
\begin{eqnarray}
N = \frac{8! \cdot 3^8}{3} \frac{12!\cdot 2^{12}}{4} = 4.33 \times 10^{19}. \label{N3}
\end{eqnarray}
The first fraction denotes the number of ways in which the eight corner cubies can be positioned ($8!$) and oriented ($3^8$) divided by the three orbits \cite{Joyner2008}, only one of which maintains the sanity \cite{Hofstadter}. The second fraction is the number of ways in which the twelve edge cubies can be placed ($12!$) and oriented ($2^{12}$) divided by the four orbits \cite{Joyner2008}, only one of which has the sanity \cite{Hofstadter}. The expected number of random configurations to generate all $N$ configurations is therefore 
\begin{eqnarray}
E[T_N] \pm \sigma[T_N] = \Big( 45.8 + 1.3\Big) \,N,
\end{eqnarray}
as per Eqs.\ (\ref{coupon}) and (\ref{sigma}).

\subsection{Half-turn metric}

In the first step of the half-turn metric, 
the 3$\times$3$\times$3  Cube can undergo any one of the following eighteen (18) turns:\ clockwise 90$^\circ$ face layer turns $R$, $D$, $B$, left ($L$), upward ($U$), front ($F$), and their inverses $R^{-1}$, $D^{-1}$, $B^{-1}$, $L^{-1}$, $U^{-1}$, $F^{-1}$  as well as 180$^\circ$ turns $R^2$, $D^2$, $B^2$, $L^2$, $U^2$, $F^2$ in the Singmaster notation \cite{Singmasterbook}.
In the second and subsequent steps, $X$, $X^{-1}$, and $X^2$ immediately following the same (where $X=R$, $D$, $B$, $L$, $U$, or $F$)
only produce the configurations that are already visited. 
Furthermore, repeated commutative turns such as $RL$ and $LR$ lead to the same configurations, together indicating $r < 15$.

With explicit evaluation \cite{Rokicki2013}, we find
\begin{eqnarray}
 S(0) = C(0) &=& 1, \label{s0} \\
 S(1) = C(1) &=& 18, \\
 S(2) = C(2) &=& 243,\\
 S(3) = C(3) &=& 3240, \label{s3}
\end{eqnarray}
suggesting $r = 3240/243 = 13.33$. Table \ref{table:3x3x3half} lists $S(t)$, $C(t)$, and $T(t)$ as a function of $t$.
For small $t$, $S(t)$ and $C(t)$ coincide with the ``positions'' listed in Table 5.1 of Ref.\ \cite{Rokicki2013}, but they are distinct quantities and deviate from each other
as $t$ increases. 
At $t=22$, $T(t)$ exceeds $E[T_N]$ for the first time and by a wide margin and, therefore,
22 is the predicted value of the diameter, which overestimates the correct value of 20 (Ref.\ \cite{Rokicki2013}) by  two.

\begin{table}
\caption{\label{table:3x3x3half} The probabilistic estimates of the numbers of non-redundant seed configurations $S(t)$,  redundant configurations $C(t)$,
and cumulative redundant configurations $T(t)$
as a function of step ($t$) in the half-turn metric of the 3$\times$3$\times$3 Cube. 
$N = 4.33\times10^{19}$, $E[T_N]   = 45.8 \,N$, and the predicted diameter is 22 for $T(22) > E[T_N] > T(21)$, while the correct diameter is 20.}
\begin{ruledtabular}
\begin{tabular}{rrrr}
$t$ & $S(t)$  & $C(t)$ & $T(t)$ \\ \hline  
0 & 1 & 1 & 1 \\
1 & 18 & 18 & 19 \\
2 & 243 & 243 & 262 \\
3 & 3240 & 3240 & 3502 \\
4 & 43217 & 43189 & 46691 \\
5 & 576232 & 576088 & 622780 \\
6 & 7683099 & 7681178 & 8303957 \\
7 & 102415704 & 102415704 & 110719662 \\
8 & 1365200187 & 1365201339 & 1475921001 \\
9 & 1.820$\times10^{10}$ & 1.820$\times10^{10}$ & 1.967$\times10^{10}$ \\
10 & 2.426$\times10^{11}$ & 2.426$\times10^{11}$ & 2.623$\times10^{11}$ \\
11 & 3.234$\times10^{12}$ & 3.234$\times10^{12}$ & 3.496$\times10^{12}$ \\
12 & 4.310$\times10^{13}$ & 4.310$\times10^{13}$ & 4.660$\times10^{13}$ \\
13 & 5.746$\times10^{14}$ & 5.746$\times10^{14}$ & 6.212$\times10^{14}$ \\
14 & 0.0002\,$N$ & 0.0002\,$N$ & 0.0002\,$N$ \\
15 & 0.0024\,$N$ & 0.0024\,$N$ & 0.0026\,$N$ \\
16 & 0.0309\,$N$ & 0.0314\,$N$ & 0.0340\,$N$ \\
17 & 0.3379\,$N$ & 0.4124\,$N$ & 0.4464\,$N$ \\
18 & 0.9889\,$N$ & 4.5046\,$N$ & 4.9510\,$N$ \\
19 & 1.0000\,$N$ & 13.1826\,$N$ & 18.1336\,$N$ \\
20 & 1.0000\,$N$ & 13.3300\,$N$ & 31.4635\,$N$ \\
21 & 1.0000\,$N$ & 13.3300\,$N$ & 44.7935\,$N$ \\
22 & 1.0000\,$N$ & 13.3300\,$N$ & 58.1235\,$N$ \\
\end{tabular}
\end{ruledtabular} 
\end{table}

Since $T(21)=44.8\,N$ is within  $\sigma[T_N]$ of $E[T_N]=45.8\,N$, Eq.\ (\ref{siobhan}) places about a 22\% chance that the diameter is actually 21, still 
overestimating the correct diameter by one.
\begin{eqnarray}
\text{Pr}(T_N \leq T(21))&=& 0.218, \\
\text{Pr}(T_N \leq T(22)) &=& 0.999998.
\end{eqnarray}
The chance that the predicted diameter agrees with the correct value of 20 is zero, which underscores the limitation of the present probabilistic estimation. 
The expected numbers of remaining unrealized configurations are
\begin{eqnarray}
N e^{-T(21)/N} &=& 1.5, \label{eq:1.5} \\
N e^{-T(22)/N} &=& 2 \times 10^{-6}.
\end{eqnarray}

\subsection{Quarter-turn metric}

In this metric, twelve (12) 90$^\circ$ turns  $R$, $D$, $B$, $L$, $U$, $F$, $R^{-1}$, $D^{-1}$, $B^{-1}$, $L^{-1}$,  $U^{-1}$, and $F^{-1}$  are valid. 
With an explicit evaluation, we find
\begin{eqnarray}
S(0) = C(0) &=& 1, \\
S(1) = C(1) &=& 12, \\
S(2) = C(2) &=& 114,\\
S(3) = C(3) &=& 1068,
\end{eqnarray}
suggesting $r = 1068/114 = 9.37$. Table \ref{table:3x3x3quarter} predicts 
how these numbers grow in subsequent steps. It indicates that at $t=26$, $T(t)$ is expected to 
exceed $E[T_N]$ for the first time and hence the predicted diameter is 26, which agrees with the correct diameter determined by Rokicki {\it et al.}\ \cite{Rokicki2013}.

\begin{table}
\caption{\label{table:3x3x3quarter} Same as Table \ref{table:3x3x3half} but in the quarter-turn metric.
The predicted diameter is 26 for $T(26) > E[T_N] > T(25)$, while the correct diameter is also 26.}
\begin{ruledtabular}
\begin{tabular}{rrrr}
$t$ & $S(t)$  & $C(t)$ & $T(t)$ \\ \hline  
0 & 1 & 1 & 1 \\
1 & 12 & 12 & 13 \\
2 & 114 & 114 & 127 \\
3 & 1068 & 1068 & 1195 \\
4 & 9604 & 10007 & 11202 \\
5 & 91237 & 89988 & 101190 \\
6 & 854745 & 854889 & 956079 \\
7 & 8009630 & 8008958 & 8965037 \\
8 & 75049468 & 75050236 & 84015273 \\
9 & 703214807 & 703213511 & 787228784 \\
10 & 6589121400 & 6589122744 & 7376351528 \\
11 & 6.174$\times10^{10}$ & 6.174$\times10^{10}$ & 6.912$\times10^{10}$ \\
12 & 5.785$\times10^{11}$ & 5.785$\times10^{11}$ & 6.476$\times10^{11}$ \\
13 & 5.421$\times10^{12}$ & 5.421$\times10^{12}$ & 6.068$\times10^{12}$ \\
14 & 5.079$\times10^{13}$ & 5.079$\times10^{13}$ & 5.686$\times10^{13}$ \\
15 & 4.759$\times10^{14}$ & 4.759$\times10^{14}$ & 5.328$\times10^{14}$ \\
16 & 0.0001\,$N$ & 0.0001\,$N$ & 0.0001\,$N$ \\
17 & 0.0010\,$N$ & 0.0010\,$N$ & 0.0011\,$N$ \\
18 & 0.0090\,$N$ & 0.0090\,$N$ & 0.0101\,$N$ \\
19 & 0.0809\,$N$ & 0.0844\,$N$ & 0.0945\,$N$ \\
20 & 0.5315\,$N$ & 0.7583\,$N$ & 0.8528\,$N$ \\
21 & 0.9931\,$N$ & 4.9804\,$N$ & 5.8332\,$N$ \\
22 & 0.9999\,$N$ & 9.3056\,$N$ & 15.1388\,$N$ \\
23 & 0.9999\,$N$ & 9.3691\,$N$ & 24.5079\,$N$ \\
24 & 0.9999\,$N$ & 9.3692\,$N$ & 33.8771\,$N$ \\
25 & 0.9999\,$N$ & 9.3692\,$N$ & 43.2463\,$N$ \\
26 & 0.9999\,$N$ & 9.3692\,$N$ & 52.6155\,$N$ \\
\end{tabular}
\end{ruledtabular} 
\end{table}

The probability that the diameter is actually 25 is less than 0.1\%; there is a 99.9\% probability that the diameter is correctly 26  according to Eq.\ (\ref{siobhan}). 
\begin{eqnarray}
\text{Pr}(T_N \leq T(25))&=& 8 \times 10^{-4}, \\
\text{Pr}(T_N \leq T(26)) &=& 0.9994.
\end{eqnarray}
The estimated numbers of configurations that have not been realized after $t=25$ and 26 are, respectively, 
\begin{eqnarray}
N e^{-T(25)/N} &=& 7.2, \\
N e^{-T(26)/N} &=& 6 \times 10^{-4}.
\end{eqnarray}
Therefore, for the quarter-turn metric, just as in the case of the 2$\times$2$\times$2 Cube, the probabilistic estimate proves to be exact. 

\subsection{Square-turn metric}

The square-turn metric \cite{Singmasterbook,Joyner2008} allows six (6) $180^\circ$ turns $R^2$, $D^2$, $B^2$, $L^2$, $U^2$, and $F^2$. 
The number of configurations in this metric is much fewer \cite{Joyner2008}.
\begin{eqnarray}
N ={4!4} \frac{4!4!4!}{2} = 6.63\times10^5. \label{N3SQ}
\end{eqnarray}
The first factor of $4!$ is the number of ways in which the front-up-right ({\it fur}) corner cubie \cite{Singmasterbook} and its three diagonal ones ({\it fdl}, {\it bdr}, {\it bul}) 
can interchange their positions by the $R^2$, $D^2$, and $B^2$ turns (their orientations are fixed by their positions).
The second factor of $4$ is the number of positions the front-up-left {\it ful} cubie can reach by the $L^2$, $U^2$, and $F^2$ turns. The remaining three 
corner cubies' positions and orientations are uniquely determined by the first five. 
The quotient $4!4!4!/2$ is the number of ways three groups of 4 edge cubies (sharing a common
center slice) can be positioned with their orientations fixed. Since the number of correctly oriented edge cubies is always even, there are two orbits
and the number must be divided by two. 

The expected number of random configurations to be generated to reach the diameter is
\begin{eqnarray}
E[T_N] \pm \sigma[T_N] = \Big( 14.0 + 1.3\Big) \,N.
\end{eqnarray}

This group has a striking similarity with the 2$\times$2$\times$2 Cube group in the quarter-turn metric (Sec.\ \ref{sec:2x2x2Q}) with the only difference being 
the number of configurations. Their $S(t)$ and $C(t)$ therefore display the identical behavior for lower $t$. 
\begin{eqnarray}
S(0) = C(0) &=& 1, \\
S(1) = C(1) &=& 6, \\
S(2) = C(2) &=& 27,\\
S(3) = C(3) &=& 120.
\end{eqnarray}
The branching ratio ($r = 120/27 = 4.44$) is also the same. 

\begin{table}
\caption{\label{table:3x3x3square}  The probabilistic estimates of the numbers of non-redundant seed configurations $S(t)$,  redundant configurations $C(t)$,
and cumulative redundant configurations $T(t)$
as a function of step ($t$) in the square-turn metric of the 3$\times$3$\times$3 Cube. 
$N = 6.63\times10^{5}$, $E[T_N]   = 14.0 \,N$.
The predicted diameter is 13 for $T(13) > E[T_N] > T(12)$, while the correct diameter is 15.}
\begin{ruledtabular}
\begin{tabular}{rrrr}
$t$ & $S(t)$  & $C(t)$ & $T(t)$ \\ \hline  
0 &	1 &	1 &	1 \\
1 &	6 &	6 &	7 \\
2 &	27 &	27 &	34 \\
3 &	120 &	120 &	154 \\
4 &	0.0008\,$N$ &	0.0008\,$N$ &	0.0010\,$N$ \\
5 &	0.0036\,$N$ &	0.0036\,$N$ &	0.0046\,$N$ \\
6 &	0.0157\,$N$ &	0.0158\,$N$ &	0.0204\,$N$ \\
7 &	0.0672\,$N$ &	0.0696\,$N$ &	0.0900\,$N$ \\
8 &	0.2580\,$N$ &	0.2984\,$N$ &	0.3884\,$N$ \\
9 &	0.6820\,$N$ &	1.1456\,$N$ &	1.5340\,$N$ \\
10 &	0.9516\,$N$ &	3.0279\,$N$ &	4.5618\,$N$ \\
11 &	0.9854\,$N$ &	4.2250\,$N$ &	8.7869\,$N$ \\
12 &	0.9874\,$N$ &	4.3751\,$N$ &	13.1619\,$N$ \\
13 &	0.9875\,$N$ &	4.3841\,$N$ &	17.5460\,$N$ \\
14 &	0.9875\,$N$ &	4.3846\,$N$ &	21.9307\,$N$ \\
15 &	0.9875\,$N$ &	4.3846\,$N$ &	26.3153\,$N$ \\
\end{tabular}
\end{ruledtabular} 
\end{table}

Table \ref{table:3x3x3square} estimates the subsequent growth and decay of 
these numbers. It can be seen that $T(t)$ exceeds $E[T_N]$ for the first time at $t=13$. This predicted diameter (13) is smaller by as much as two than
the correct diameter of 15, which has been determined by brute-force computation. Worse yet, as per Eq.\ (\ref{siobhan}), the probability that the
predicted diameter is actually 12 instead of 13 is not negligible.
\begin{eqnarray}
\text{Pr}(T_N \leq T(12)) &=& 0.279, \\
\text{Pr}(T_N \leq T(13)) &=& 0.984.
\end{eqnarray}
In other words, there is a fair chance that one last configuration is yet to be realized after 12 steps. 
\begin{eqnarray}
N e^{-T(12)/N} &=& 1.3, \\
N e^{-T(13)/N} &=& 0.016,
\end{eqnarray}
After 13 steps, all configurations should have been generated according to our estimates. Hence, this metric turns out to be the hardest case 
for the probabilistic estimates.

%==============================
\section{The 4$\times$4$\times$4 Cube}
%==============================

The number of configurations of the 4$\times$4$\times$4 Cube is \cite{singmaster4x4x4}
\begin{eqnarray}
N = \frac{7! \cdot 3^7}{3} 24! \frac{ 24!}{(4!)^6} = 7.40 \times 10^{45}. \label{N4}
\end{eqnarray}
The first fraction is the number of ways in which seven of the eight corner cubies can be placed and oriented, while the last (eighth) one is held fixed as the orientation anchor. 
This fraction is therefore equal to the number of configurations of the 2$\times$2$\times$2 Cube [Eq.\ (\ref{N2})]. 
The second factor ($24!$) is the number of ways in which twenty-four edge cubies can be permuted, which are distinguishable and 
cannot be freely oriented. The last fraction is the number of ways in which six quadruplicate center cubies can be positioned, whose orientations
cannot be freely changed. 

The number of random configurations that needs to be generated before all $N$ configurations occur is 
\begin{eqnarray}
E[T_N] \pm \sigma[T_N] =\Big( 106.2 \pm 1.3 \Big)\,N.
\end{eqnarray}
according to Eqs.\ (\ref{coupon}) and (\ref{sigma}). 

\subsection{Half-turn metric}

There are twenty-seven (27) turns, which are clockwise 90$^\circ$  turns  $R_1$, $D_1$, $B_1$, $R_2$, $D_2$, $B_2$, $R_3$, $D_3$, $B_3$ and their inverses 
$R_1^{-1}$, $D_1^{-1}$, $B_1^{-1}$, etc.\ as well as
the corresponding 180$^\circ$ turns $R_1^{2}$, $D_1^2$, $B_1^2$, etc., where
the subscript $i$ in $X_i$ refers to the $i$th layer rotated 
with $i=1$ being the face (exterior) layer (and hence $R_3$ is alternatively denoted by $L_2^{-1}$).

For the first few steps, we can explicitly compute
\begin{eqnarray}
S(0) = C(0) &=& 1, \\
S(1) = C(1) &=& 27, \\
S(2) = C(2) &=& 567,\\
S(3) = C(3) &=& 11721,
\end{eqnarray}
and, therefore, we adopt $r = 11721/567 = 20.67$. Table \ref{table:4x4x4half} estimates $S(t)$, $C(t)$, and $T(t)$ for $t \geq 4$ using Eqs.\ (\ref{growth})--(\ref{G}).
 At $t=41$, $T(t)$ becomes greater than $E[T_N]$ for the first time, 
predicting the diameter of 41 for the 4$\times$4$\times$4 Cube in the half-turn metric. The correct diameter 
is unknown and its brute-force computational determination seems out of the question in the foreseeable future. 

\begin{table}
\caption{\label{table:4x4x4half} The probabilistic estimates of the non-redundant seed configurations $S(t)$,  redundant configurations $C(t)$,
and cumulative redundant configurations $T(t)$
as a function of step ($t$) in the half-turn metric of the 4$\times$4$\times$4 Cube.
$N = 7.40\times10^{45}$, $E[T_N]  =106.2 \,N$, and the predicted diameter is 41 since $T(41) > E[T_N] > T(40)$.
The correct diameter is unknown.}
\begin{ruledtabular}
\begin{tabular}{rrrr}
$t$ & $S(t)$  & $C(t)$ & $T(t)$ \\ \hline  
0 &	1 &	1 &	1 \\
1 &	27 &	27 &	28 \\
2 &	567 &	567 &	595 \\
3 &	11721 &	11721 &	12316 \\
4 &	242273 &	242273 &	254589 \\
5 &	5007784 &	5007784 &	5262373 \\
6 &	103510903 &	103510903 &	108773276 \\
7 &	2139570358 &	2139570358 &	2248343634 \\
8 &	44224919298 &	44224919298 &	46473262932 \\
$\cdots$  & $\cdots$  & $\cdots$  & $\cdots$  \\
%9 &	9.141E+11 &	9.141E+11 &	9.606E+11 \\
%10 &	1.890E+13 &	1.890E+13 &	1.986E+13 \\
%11 &	3.906E+14 &	3.906E+14 &	4.104E+14 \\
%12 &	8.073E+15 &	8.073E+15 &	8.483E+15 \\
%13 &	1.669E+17 &	1.669E+17 &	1.753E+17 \\
%14 &	3.449E+18 &	3.449E+18 &	3.624E+18 \\
%15 &	7.129E+19 &	7.129E+19 &	7.492E+19 \\
%16 &	1.474E+21 &	1.474E+21 &	1.549E+21 \\
%17 &	3.046E+22 &	3.046E+22 &	3.201E+22 \\
%18 &	6.296E+23 &	6.296E+23 &	6.616E+23 \\
%19 &	1.301E+25 &	1.301E+25 &	1.368E+25 \\
%20 &	2.690E+26 &	2.690E+26 &	2.827E+26 \\
%21 &	5.560E+27 &	5.560E+27 &	5.843E+27 \\
%22 &	1.149E+29 &	1.149E+29 &	1.208E+29 \\
%23 &	2.376E+30 &	2.376E+30 &	2.496E+30 \\
%24 &	4.910E+31 &	4.910E+31 &	5.160E+31 \\
%25 &	1.015E+33 &	1.015E+33 &	1.067E+33 \\
%26 &	2.098E+34 &	2.098E+34 &	2.205E+34 \\
%27 &	4.336E+35 &	4.336E+35 &	4.557E+35 \\
%28 &	8.963E+36 &	8.963E+36 &	9.419E+36 \\
%29 &	1.853E+38 &	1.853E+38 &	1.947E+38 \\
%30 &	3.830E+39 &	3.830E+39 &	4.024E+39 \\
%31 &	7.916E+40 &	7.916E+40 &	8.318E+40 \\
32 &	0.0002\,$N$ &	0.0002\,$N$ &	0.0002\,$N$ \\
33 &	0.0046\,$N$ &	0.0046\,$N$ &	0.0048\,$N$ \\
34 &	0.0899\,$N$ &	0.0942\,$N$ &	0.0990\,$N$ \\
35 &	0.8442\,$N$ &	1.8590\,$N$ &	1.9581\,$N$ \\
36 &	1.0000\,$N$ &	17.4491\,$N$ &	19.4072\,$N$ \\
37 &	1.0000\,$N$ &	20.6700\,$N$ &	40.0772\,$N$ \\
38 &	1.0000\,$N$ &	20.6700\,$N$ &	60.7472\,$N$ \\
39 &	1.0000\,$N$ &	20.6700\,$N$ &	81.4172\,$N$ \\
40 &	1.0000\,$N$ &	20.6700\,$N$ &	102.0872\,$N$ \\
41 &	1.0000\,$N$ &	20.6700\,$N$ &	122.7572\,$N$ \\
\end{tabular}
\end{ruledtabular} 
\end{table}

$T(41)$ exceeds $E[T_N]$ by a wide margin and hence the probability that the diameter is actually 40 is  zero.
\begin{eqnarray}
\text{Pr}(T_N \leq T(40))&=& 1 \times 10^{-15}, \\
\text{Pr}(T_N \leq T(41)) &=& 1.
\end{eqnarray}
However, these probabilities are predicated on the validity of the assumptions underlying our probabilistic predictions, which may well overestimate
the diameters judging from the results for the 2$\times$2$\times$2 and 3$\times$3$\times$3 Cubes in the half-turn metric. 
The estimated numbers of unrealized configurations after $t=40$ and $41$ are
\begin{eqnarray}
N e^{-T(40)/N} &=& 34, \\
N e^{-T(41)/N} &=& 4 \times 10^{-8},
\end{eqnarray}
reiterating the fact that the present probabilistic estimate of 41 is rather definitive because the stepwise growth of $T(t)$ for large $t$ is $20.7\,N$, which is an order of magnitude 
greater than $\sigma[T_N] = 1.3\,N$. 

\subsection{Quarter-turn metric}

There are eighteen (18) turns in this metric, 
which are clockwise 90$^\circ$  turns  $R_1$, $D_1$, $B_1$, $R_2$, $D_2$, $B_2$,  and $R_3$, $D_3$, $B_3$  as well as their inverses.

An explicit enumeration shows 
\begin{eqnarray}
S(0) = C(0) &=& 1, \\
S(1) = C(1) &=& 18, \\
S(2) = C(2) &=& 261,\\
S(3) = C(3) &=& 3732.
\end{eqnarray}
We adopt $r = 3732/261 = 14.30$. Using the identical probabilistic logic in the foregoing sections,
we obtain Table \ref{table:4x4x4quarter}.
At $t=48$, $T(t)$ exceeds $E[T_N]$ for the first time, and therefore the predicted diameter is 48, whereas the correct value is unknown.

\begin{table}
\caption{\label{table:4x4x4quarter} Same as Table \ref{table:4x4x4half} but in the quarter-turn metric. 
The predicted diameter is 48 since $T(48) > E[T_N] > T(47)$.
The correct diameter is unknown.}
\begin{ruledtabular}
\begin{tabular}{rrrr}
$t$ & $S(t)$  & $C(t)$ & $T(t)$ \\ \hline  
0 &	1 & 	1 & 	1 \\
1 &	18 &	18 &	19 \\
2 &	261 &	261 &	280 \\
3 &	3732 &	3732 &	4012 \\
4 &	53368 &	53368 &	57380 \\
5 &	763157 &	763157 &	820536 \\
6 &	10913141 &	10913141 &	11733677 \\
7 &	156057909 &	156057909 &	167791586 \\
8 &	2231628106 &	2231628106 &	2399419692 \\
9 &	31912281912 &	31912281912 &	34311701604 \\
$\cdots$  & $\cdots$  & $\cdots$  & $\cdots$  \\
36 &	0.0001\,$N$ &	0.0001\,$N$ &	0.0001\,$N$ \\
37 &	0.0010\,$N$ &	0.0010\,$N$ &	0.0010\,$N$ \\
38 &	0.0137\,$N$ &	0.0138\,$N$ &	0.0148\,$N$ \\
39 &	0.1777\,$N$ &	0.1957\,$N$ &	0.2105\,$N$ \\
40 &	0.9213\,$N$ &	2.5417\,$N$ &	2.7522\,$N$ \\
41 &	1.0000\,$N$ &	13.1741\,$N$ &	15.9264\,$N$ \\
42 &	1.0000\,$N$ &	14.3000\,$N$ &	30.2264\,$N$ \\
43 &	1.0000\,$N$ &	14.3000\,$N$ &	44.5263\,$N$ \\
44 &	1.0000\,$N$ &	14.3000\,$N$ &	58.8263\,$N$ \\
45 &	1.0000\,$N$ &	14.3000\,$N$ &	73.1263\,$N$ \\
46 &	1.0000\,$N$ &	14.3000\,$N$ &	87.4263\,$N$ \\
47 &	1.0000\,$N$ &	14.3000\,$N$ &	101.7263\,$N$ \\
48 &	1.0000\,$N$ &	14.3000\,$N$ &  116.0263\,$N$ \\
\end{tabular}
\end{ruledtabular} 
\end{table}

The probabilities that the diameter is 47 or 48 are, respectively,
\begin{eqnarray}
\text{Pr}(T_N \leq T(47))&=& 5 \times 10^{-22}, \\
\text{Pr}(T_N \leq T(48)) &=& 0.99997.
\end{eqnarray}
The estimated numbers of unrealized configurations are
\begin{eqnarray}
N e^{-T(47)/N} &=& 49, \\
N e^{-T(48)/N} &=& 3 \times 10^{-5},
\end{eqnarray}
meaning that the predicted diameter of 48 is definitive, which however says nothing about the accuracy of the prediction.

%==============================
\section{The 5$\times$5$\times$5 Cube}
%==============================

The number of configurations of the 5$\times$5$\times$5 Cube is \cite{singmaster4x4x4}
\begin{eqnarray}
N = \frac{8! \cdot 3^8 }{3} \frac{12! \cdot 2^{12}}{4}  24!  \frac{ 24!}{(4!)^6} \frac{ 24!}{(4!)^6}= 2.83 \times 10^{74}. \label{N5}
\end{eqnarray}
The center cubie of each face is held fixed as the orientation anchor. 
The first two fractions are, respectively, the number of ways in which the eight corner cubies are placed and oriented and the number of ways in which 
the twelve center-edge cubies are positioned and oriented. 
They are the same as the corresponding factors for the 3$\times$3$\times$3 Cube [Eq.\ (\ref{N3})]. The third factor ($24!$) is the number of ways in which twenty-four non-central-edge cubies
are permuted, which are distinguishable and cannot be freely oriented. The last two fractions are, respectively, the number of ways in which six quadruplicate 
diagonal center cubies are positioned and the number of ways in which six quadruplicate remaining cubies are placed. 

Therefore, the expected number of configurations that need to be generated before every configuration is realized is 
\begin{eqnarray}
E[T_N] \pm \sigma[T_N] =\Big( 172.0 \pm 1.3 \Big)\,N.
\end{eqnarray}

\subsection{Half-turn metric}

The thirty-six (36) turns in this metric are face (exterior) layer clockwise 90$^\circ$  turns $R_1$, $D_1$, $B_1$,   $L_1$, $U_1$, $F_1$,
interior layer clockwise 90$^\circ$  turns  $R_2$, $D_2$, $B_2$, $L_2$, $U_2$, $F_2$,  and their inverses as well as the corresponding 180$^\circ$ turns. 

In the first few steps, the numbers of configurations can be determined exactly as
\begin{eqnarray}
S(0) = C(0) &=& 1, \\
S(1) = C(1) &=& 36, \\
S(2) = C(2) &=& 1026,\\
S(3) = C(3) &=& 28812,
\end{eqnarray}
suggesting $r = 28812/1026 = 28.08$. In the subsequent steps, they are estimated probabilistically, and the result is given in Table \ref{table:5x5x5half}.
It shows that $T(58)$ is greater than $E[T_N]$ for the first time and hence the predicted diameter is 58. The correct value of the diameter is unknown. 

\begin{table}
\caption{\label{table:5x5x5half} The probabilistic estimates of the non-redundant seed configurations $S(t)$,  redundant configurations $C(t)$,
and cumulative redundant configurations $T(t)$
as a function of step ($t$) in the quarter-turn metric of the 5$\times$5$\times$5 Cube.
$N = 2.83\times10^{74}$, $E[T_N] =172.0\,N$, and the predicted diameter is 58 since $T(58) > E[T_N] > T(57)$.
The correct diameter is unknown.}
\begin{ruledtabular}
\begin{tabular}{rrrr}
$t$ & $S(t)$  & $C(t)$ & $T(t)$ \\ \hline  
0 &	1 &	1 &	1 \\
1 &	36 &	36 &	37 \\
2 &	1026 &	1026 &	1063 \\
3 &	28812 &	28812 &	29875 \\
4 &	809041 &	809041 &	838916 \\
5 &	22717870 &	22717870 &	23556786 \\
6 &	637917794 &	637917794 &	661474580 \\
7 &	17912731656 &	17912731656 &	18574206236 \\
$\cdots$ & $\cdots$ & $\cdots$ & $\cdots$ \\
49 &	0.0004\,$N$ &	0.0004\,$N$ &	0.0004\,$N$ \\
50 &	0.0120\,$N$ &	0.0121\,$N$ &	0.0125\,$N$ \\
51 &	0.2864\,$N$ &	0.3374\,$N$ &	0.3499\,$N$ \\
52 &	0.9997\,$N$ &	8.0413\,$N$ &	8.3912\,$N$ \\
53 &	1.0000\,$N$ &	28.0710\,$N$ &	36.4621\,$N$ \\
54 &	1.0000\,$N$ &	28.0800\,$N$ &	64.5421\,$N$ \\
55 &	1.0000\,$N$ &	28.0800\,$N$ &	92.6221\,$N$ \\
56 &	1.0000\,$N$ &	28.0800\,$N$ &	120.7021\,$N$ \\
57 &	1.0000\,$N$ &	28.0800\,$N$ &	148.7821\,$N$ \\
58 &	1.0000\,$N$ &	28.0800\,$N$ &	176.8621\,$N$ \\
\end{tabular}
\end{ruledtabular} 
\end{table}

Owing to the large number of configurations ($28.1\,N$) generated in each of the later steps as compared with the fixed standard deviation of $\sigma[T_N] = 1.3\,N$, 
the probability that the predicted diameter is shifted by one is essentially zero.
\begin{eqnarray}
\text{Pr}(T_N \leq T(57))&=& 0, \\
\text{Pr}(T_N \leq T(58)) &=& 0.996.
\end{eqnarray}
Correspondingly, many configurations are expected to remain unrealized after $t=57$, all of which are however recuperated in the next step.
\begin{eqnarray}
N e^{-T(57)/N} &=& 7 \times 10^9, \\
N e^{-T(58)/N} &=& 0.004.
\end{eqnarray}
Therefore, the present probabilistic estimate is definitive, saying nothing about its accuracy.

\subsection{Quarter-turn metric}

The twenty-four (24) turns in this metric are exterior-layer clockwise 90$^\circ$  turns, 
interior-layer clockwise 90$^\circ$  turns,  and their inverses. An explicit enumeration gives
\begin{eqnarray}
S(0) = C(0) &=& 1, \\
S(1) = C(1) &=& 24, \\
S(2) = C(2) &=& 468,\\
S(3) = C(3) &=& 9000,
\end{eqnarray}
suggesting $r = 9000/468 = 19.23$. Table \ref{table:5x5x5quarter} indicates that the predicted diameter is 68 because $T(68) > E[T_N] > T(67)$. 

\begin{table}
\caption{\label{table:5x5x5quarter} Same as Table \ref{table:5x5x5half} but in the quarter-turn metric. The predicted diameter is 68 since $T(68) > E[T_N] > T(67)$.
The correct diameter is unknown.}
\begin{ruledtabular}
\begin{tabular}{rrrr}
$t$ & $S(t)$  & $C(t)$ & $T(t)$ \\ \hline  
0 &	1 &	1 &	1 \\
1 &	24 &	24 &	25 \\
2 &	468 &	468 &	493 \\
3 &	9000 &	9000 &	9493 \\
4 &	173070 &	173070 &	182563 \\
5 &	3328136 &	3328136 &	3510699 \\
6 &	64000057 &	64000057 &	67510756 \\
7 &	1230721100 &	1230721100 &	1298231856 \\
8 &	23666766753 &	23666766753 &	24964998610 \\
9 &	455111924665 &	455111924665 &	480076923275 \\
$\cdots$ & $\cdots$ & $\cdots$ & $\cdots$ \\
55 &	0.0002\,$N$ &	0.0002\,$N$ &	0.0002\,$N$ \\
56 &	0.0036\,$N$ &	0.0036\,$N$ &	0.0038\,$N$ \\
57 &	0.0663\,$N$ &	0.0686\,$N$ &	0.0724\,$N$ \\
58 &	0.7207\,$N$ &	1.2755\,$N$ &	1.3479\,$N$ \\
59 &	1.0000\,$N$ &	13.8593\,$N$ &	15.2072\,$N$ \\
60 &	1.0000\,$N$ &	19.2300\,$N$ &	34.4371\,$N$ \\
61 &	1.0000\,$N$ &	19.2300\,$N$ &	53.6671\,$N$ \\
62 &	1.0000\,$N$ &	19.2300\,$N$ &	72.8971\,$N$ \\
63 &	1.0000\,$N$ &	19.2300\,$N$ &	92.1271\,$N$ \\
64 &	1.0000\,$N$ &	19.2300\,$N$ &	111.3571\,$N$ \\
65 &	1.0000\,$N$ &	19.2300\,$N$ &	130.5871\,$N$ \\
66 &	1.0000\,$N$ &	19.2300\,$N$ &	149.8171\,$N$ \\
67 &	1.0000\,$N$ &	19.2300\,$N$ &	169.0471\,$N$ \\
68 &	1.0000\,$N$ &	19.2300\,$N$ &	188.2771\,$N$ \\
%69 &	1.0000\,$N$ &	19.2300\,$N$ &	207.5071\,$N$ \\
\end{tabular}
\end{ruledtabular} 
\end{table}

In this case also, the probability that the diameter is 67 instead of 68 is virtually zero.
\begin{eqnarray}
\text{Pr}(T_N \leq T(67))&=& 2 \times 10^{-5}, \\
\text{Pr}(T_N \leq T(68)) &=& 1.
\end{eqnarray}
There expected to be an appreciable number of configurations that have not been realized after $t=67$, but none at $t=68$. 
\begin{eqnarray}
N e^{-T(67)/N} &=& 11, \\
N e^{-T(68)/N} &=& 5 \times 10^{-8}.
\end{eqnarray}

\section{Discussion}

Table \ref{table:summary} summarizes the size and metric dependence of the actual (if available) and predicted
diameters of the $n\times n \times n$ Rubik's Cube groups ($ 2 \leq n \leq 5$). This table's predictions can be easily extended  to greater $n$, different metrics, or 
other types of Cubes with essentially little additional computational cost, but determining actual values will be formidable in many cases. 

\begin{table}
\caption{\label{table:summary} Actual and predicted diameters of the $n \times n \times n$ Rubik's Cube group ($2 \leq n \leq 5$) 
in various metrics.}
\begin{ruledtabular}
\begin{tabular}{llrrrr}
& && \multicolumn{3}{c}{Diameter} \\ \cline{4-6}
\multicolumn{1}{l}{Cube} & \multicolumn{1}{l}{Metric} & \multicolumn{1}{c}{$r$\footnotemark[1]}& \multicolumn{1}{c}{Actual} & \multicolumn{1}{c}{Predicted\footnotemark[2]} & \multicolumn{1}{c}{Predicted\footnotemark[3]} \\ \hline
2$\times$2$\times$2 & Half  & 5.94 & 11\footnotemark[4] &  12 & 11.0 \\ %11.0 \\
2$\times$2$\times$2 & Quarter  &4.44 & 14\footnotemark[4] &  14 & 13.5  \\ %13.5 \\
2$\times$2$\times$2 & Semi-quarter  &2.77 & 19\footnotemark[5]  &  21 & 20.3 \\ %, 19\footnotemark[7]  \\ %20.3 \\
2$\times$2$\times$2 & Bi-quarter  &9.19 & 10\footnotemark[5]  &  9 & 8.5 \\ %8.5 \\
%2$\times$2$\times$2 & Square  & 2.98 & 4 &  $\dots$ & 4\footnotemark[7]  \\ %8.5 \\
3$\times$3$\times$3 & Half  &13.33 & 20\footnotemark[6] &  22 & 20.8 \\ %20.8 \\
3$\times$3$\times$3 & Quarter & 9.37 & 26\footnotemark[6] &  26 & 25.0 \\ %25.0 \\
3$\times$3$\times$3 & Square & 4.44 & 15\footnotemark[5] &  13 & 12.0 \\ %25.0 \\
%3$\times$3$\times$3 & Square-slice & 2.89 & 3 &  $\dots$ & 3\footnotemark[7]  \\ %25.0 \\
4$\times$4$\times$4 & Half & 20.67 & $\cdots$ &  41 & 40.0 \\ %47.1 \\
4$\times$4$\times$4 & Quarter & 14.30 & $\cdots$ &  48 & 47.1 \\ %47.1 \\
5$\times$5$\times$5 & Half & 28.08 & $\cdots$ &  58 & 57.5 \\ %66.9 \\
5$\times$5$\times$5 & Quarter & 19.23 & $\cdots$ &  68 & 66.9 \\ %66.9 \\
\end{tabular}
\end{ruledtabular} 
\footnotetext[1]{Branching ratio. See Eq.\ (\ref{growth}).}
\footnotetext[2]{Probabilistic estimate based on a modified version of the coupon collector's problem (this work).}
\footnotetext[3]{Equation (\ref{asymp}) (this work).}
\footnotetext[4]{Reference \cite{rewix}.}
\footnotetext[5]{This work.}
\footnotetext[6]{Rokicki {\it et al.}\ \cite{cube20,Rokicki2013,Rokicki2014}.}
%\footnotetext[7]{Equations (\ref{approx_r}) and (\ref{approx_d}) (this work).}
\end{table}

\subsection{Comparison between predicted and actual results\label{sec:comparison}}

\begin{figure}
  \includegraphics[width=\columnwidth]{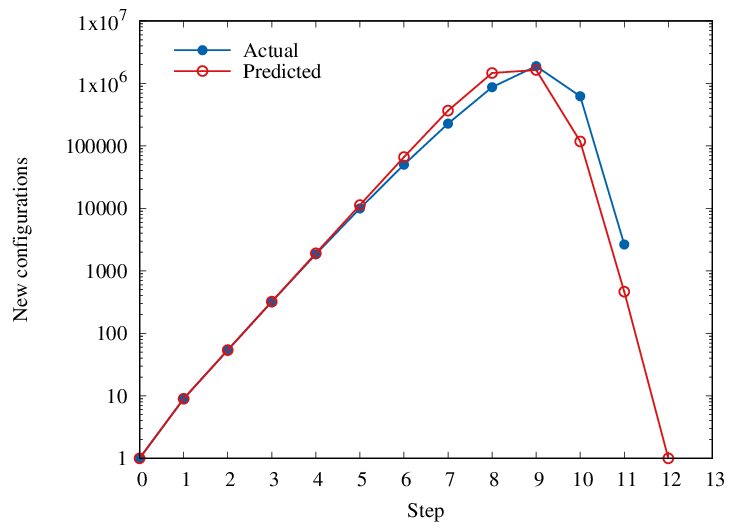}
\caption{The number of new (non-redundant) configurations of the 2$\times$2$\times$2 Cube generated in each step in the half-turn metric. A terminal plot 
(step 11 for the actual) corresponds to the diameter.}
\label{fig:2Hnews}
\end{figure}

\begin{figure}
  \includegraphics[width=\columnwidth]{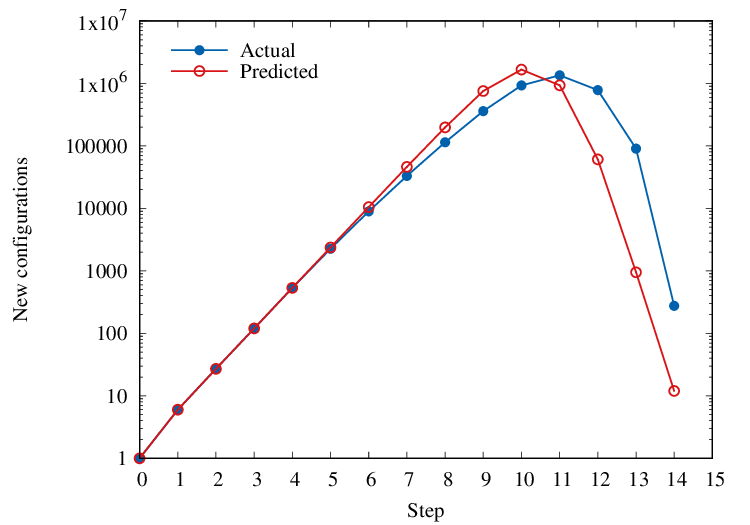}
\caption{Same as Fig.\ \ref{fig:2Hnews} for the quarter-turn metric.}
\label{fig:2Qnews}
\end{figure}

\begin{figure}
  \includegraphics[width=\columnwidth]{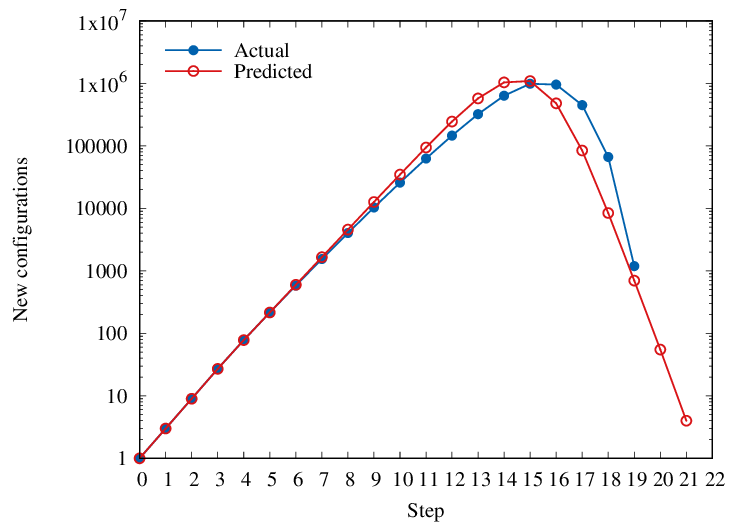}
\caption{Same as Fig.\ \ref{fig:2Hnews} for the semi-quarter-turn metric.}
\label{fig:2SemiQnews}
\end{figure}

\begin{figure}
  \includegraphics[width=\columnwidth]{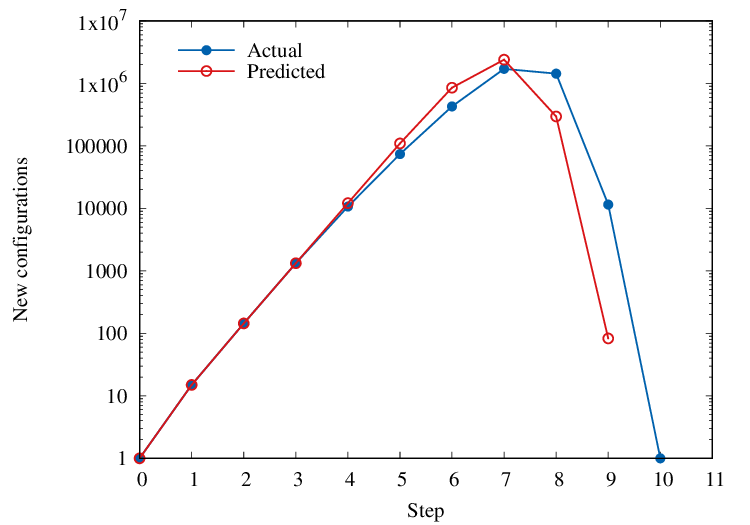}
\caption{Same as Fig.\ \ref{fig:2Hnews} for the bi-quarter-turn metric.}
\label{fig:2BiQnews}
\end{figure}

For the 2$\times$2$\times$2 Cube, one can easily perform the brute-force enumeration of configurations as a function of the number of turns applied to 
the solved configuration, in nearly but not exactly the same way in which the probabilistic estimation's hypothetical algorithm generates configurations. 

In Figs.\ \ref{fig:2Hnews}--\ref{fig:2BiQnews} are plotted the number of new, non-redundant configurations generated (``Actual'') in each step for the half-turn,
quarter-turn, semi-quarter-turn, and bi-quarter-turn metrics, respectively, which are compared with the probabilistic estimates (``Predicted''). 
The ``Actual'' numbers were determined by a computer program \cite{hirata_rubik}. The ``Predicted'' numbers were obtained as
\begin{eqnarray}
N \left( e^{-T(t-1)/N} - e^{-T(t)/N} \right),
\end{eqnarray}
which is consistent with Appendix \ref{sec:derivation2}.
A terminal point corresponds to the diameter at which all configurations have been generated and no new ones can be added in the subsequent steps. 

In each of the metrics, the actual curve in earlier steps displays a geometrical growth $r^t$ with the step count $t$ 
and is accurately reproduced by the probabilistically predicted curve. 
The peaks of the curves tend to occur one step too early in the predicted curves. 
The subsequent rapid falloffs of the actual curves seem super-exponential with their slopes in the logarithm scale becoming more negative for higher step counts.  
It may be imagined that in the final few steps of the diameter, the remaining unrealized configurations will be more efficiently exhausted than a random generation can.
However, the predicted curves, owing to the non-Markovian assumption of the probabilistic estimation, will display an exponential and thus slower decay 
of the form $e^{-rt}$, where $r$ is the branching ratio, taking more steps than reality to exhaust the remaining unrealized configurations. 

In the half-turn and quarter-turn metrics, therefore, there seems to be a systematic cancellation of errors between the peak positions  
(occurring one step too early in the predicted curves) and the rate of decay (which is faster in the actual curves), although the predicted diameter
in the half-turn metric is still overestimated by one. 
In the semi-quarter-turn metric, since $r$ is small, there is a longer decay phase, causing the slow decay of the predicted curve to overshoot
the diameter by as much as two, despite  its peak position occurring one step too early. In the bi-quarter-turn metric, in contrast, 
$r$ is large, which minimizes the error of the predicted decay rate, and underestimates the diameter by one. 

Implicit in our probabilistic estimation logic is the assumption that the details of each metric are unimportant and all of its characteristics
can be encapsulated into the branching ratio $r$, which is invariant of the step count. 
From Figs.\ \ref{fig:2Hnews}--\ref{fig:2BiQnews}, 
this assumption seems largely valid for the growth phase, but is no longer so in the decay phase. 
%The fact that these curves show a super-exponential decay suggests that the exponent factor $r$ in the $e^{-rt}$ decay of the predicted curve should increase near the diameter and may even exceed the number of turns if . In fact, the actual and predicted curves  in Fig.\ \ref{fig:2SemiQnews} behave like $e^{-4.0t}$ and $e^{-2.6t}$(where $t$ is the number of steps) at the diameter, respectively, and the factor of 4.0 in the former exceeds $r=2.77$ or even the number of turns, which is three. There seems room for improvement in the probabilistic estimates in the peak position and decay rate of these curves.

\begin{figure}
\includegraphics[width=\columnwidth]{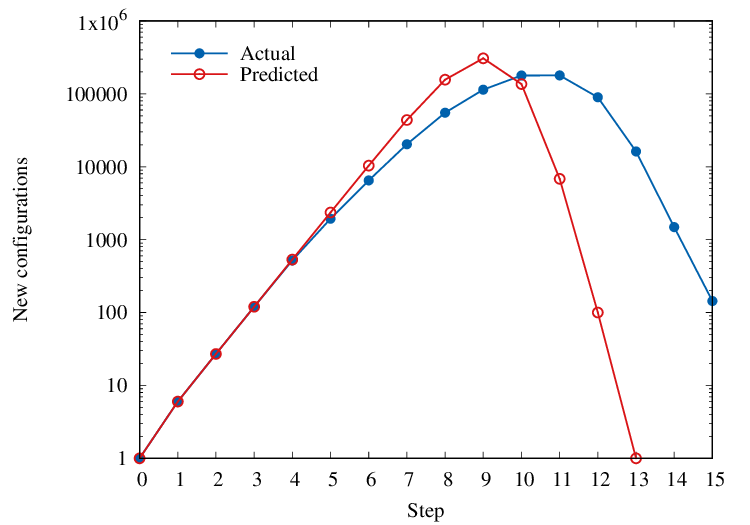}
\caption{The number of new (non-redundant) configurations of the 3$\times$3$\times$3 Cube generated in each step in the square-turn metric. }
\label{fig:3SQnews}
\end{figure}

The 3$\times$3$\times$3 Cube in the square-turn metric can be solved by brute force because its number of configurations is relatively small ($N = 6.63 \times 10^5$).
This is an interesting case: It is the same as the 2$\times$2$\times$2 Cube in the quarter-turn metric in every respect except for $N$, but the probabilistic 
estimate works most poorly for the former, while it gives an exact prediction for the latter. In fact, the diameter for the 3$\times$3$\times$3 Cube group in the square-turn metric
with a smaller $N$ is greater by one than the diameter for the 2$\times$2$\times$2 Cube group in the quarter-turn metric with a larger $N$; the order and diameter of locally isomorphic graphs are inverted \cite{Hirata2024_geom}.

Figure \ref{fig:3SQnews} attests to the large deviations in 
the actual and predicted numbers of new configurations for the 3$\times$3$\times$3 Cube in the square-turn metric.
The errors are seen already around the peaks and aggravated
in the falling part. Unlike the foregoing cases, the decay of the actual curve is unusually slow, resulting in the inversion, and slower than the predicted curve's decay. 
The cause of the slow decay is unknown, but the poor performance of the probabilistic logic may be ascribed to the relative smallness of $N$. 

\begin{figure}
  \includegraphics[width=\columnwidth]{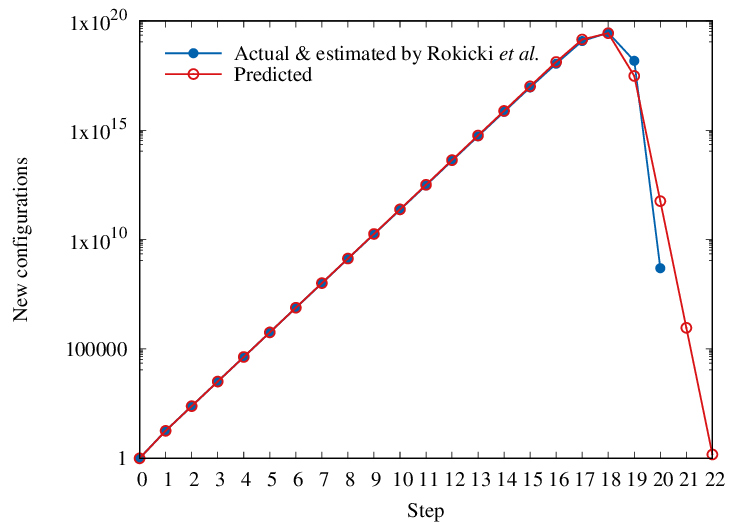}
\caption{The number of new (non-redundant) configurations of the 3$\times$3$\times$3 Cube generated in each step in the half-turn metric. 
The actual values (those in the last five steps were approximate) were taken from Rokicki {\it et al.}\ \cite{Rokicki2013,cube20}.}
\label{fig:3Hnews}
\end{figure}

\begin{figure}
  \includegraphics[width=\columnwidth]{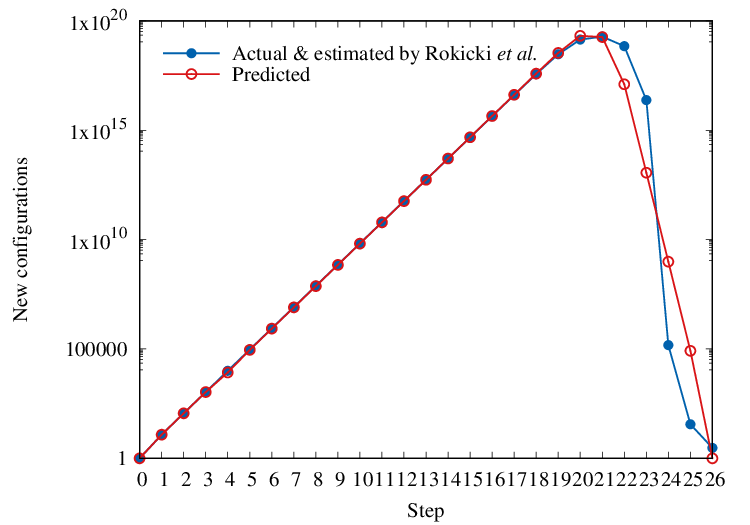}
\caption{The number of new (non-redundant) configurations of the 3$\times$3$\times$3 Cube generated in each step in the quarter-turn metric. 
The actual values (those in steps 19 through 26 were approximate) were taken from Rokicki {\it et al.}\ \cite{cube20}.}
\label{fig:3Qnews}
\end{figure}

For the 3$\times$3$\times$3 Cube in the half- and quarter-turn metrics, 
Rokicki {\it et al.}\ \cite{Rokicki2013,cube20} gave the actual and estimated numbers (``positions'') of new configurations generated 
at each step.
They are compared with our predicted values in Figs.\ \ref{fig:3Hnews} and \ref{fig:3Qnews}. 

For the half-turn metric (Fig.\ \ref{fig:3Hnews}), the rising part and the peak of the actual curve is accurately reproduced 
by the prediction, but the falling part decays too slow in the latter, overestimating the diameter by two.
The actual curve has a rapid falloff of $e^{-22.3t}$, and it likely decreases even faster than exponentially. 
On the other hand, the predicted curve decays exponentially as $e^{-13.3 t}$ corresponding to $r=13.33$.
It again suggests that our probabilistic logic tends to underestimate the accelerated rate of exhausting remaining unrealized configurations in the final few steps.

Singmaster \cite{Singmasterbook}, using the same argument, inferred that nearly all configurations should be generated 
at 17.3 steps, which is only slightly less than our value or the one of Rokicki {\it et al.}\ \cite{Rokicki2013}.
On this basis, Singmaster speculated that ``every position can be achieved in at most 20 moves,'' which 
is in exact agreement with the correct value determined by Rokicki {\it et al.}\ \cite{Rokicki2013} long before the latter became available. 
No account seems to have been given about the addition of the 3 moves, however.

For the quarter-turn metric (Fig.\ \ref{fig:3Qnews}), on the other hand, the rising part of the actual curve is accurately reproduced by the prediction, while 
the falling part of the predicted curve initially underestimates and then overestimates the rate of decay. 
This is simply because the ``actual'' curve (which is only approximate \cite{cube20} for steps 19 through 26) shows an unusual (as compared with all the other cases discussed above) or even puzzling
behavior of slowing down the rate of decay in the last three steps. The Cayley graph of this metric is expected to be a symmetric graph because every configuration is equivalent and so is every turn \cite{Hirata2024_geom}.
For such a graph, the branching ratio $r$ is a monotonically decreasing function of step count \cite{Hirata2024_geom}, which seems to be contradicted by the ``actual'' curve \cite{cube20}.

Overall, however, the agreement is good in Figs.\ \ref{fig:3Hnews} and \ref{fig:3Qnews}, supporting the notion that probabilistic logic generally works progressively well for larger $N$. 

In the 4$\times$4$\times$4 and 5$\times$5$\times$5 Cubes, whose branching ratio $r$ is small relative to the huge $E[T_N]$, there expected to be 
long decay phases, potentially causing noticeable overestimation of the diameters. At the same time, it is also possible that their huge $N$ make the probabilistic estimates
more accurate. 

\subsection{Size dependence\label{sec:size}}

Demaine {\it et al.}\ \cite{Demaine2011} argued that the diameter is asymptotically proportional to $n^2 / \log n$ for the $n\times n \times n$ Cube.
Under the assumption of the exact proportionality in the quarter-turn metric, their formula predicts the diameters of the 4$\times$4$\times$4 
and 5$\times$5$\times$5 Cubes in the same metric to be 43 and 62, respectively, which are much smaller than our estimates (48 and 68). 

Our probabilistic estimation suggests that a more meaningful closed-form expression may be  
a function of the number of configurations $N$ (rather than of the Cube size $n$) and also of the metric. This is because there are complicated $n$ dependencies
(possibly including even-odd alternation) of $N$ and also because it is $N$ that dictates the complexity of the puzzle and thus the diameters more directly  than $n$. 
Furthermore, the diameters may also depend on the metric in a nontrivial manner since some turns may be  
more useful than others. 

With the branching ratio $r$ and the number of configurations $N$, both of which can be relatively easily determined as has been done in the foregoing sections, 
the probabilistic estimate ($d$) of the diameter (which must be distinguished from the correct diameter) is approximated by
\begin{eqnarray}
d \approx \frac{\ln N}{\ln r} + \frac{\ln N}{r}.  \label{asymp}
\end{eqnarray}
The first term counts the number of steps required for $S(t)$ to increase geometrically and reach $N$, while the second term 
is the number of additional steps needed for $T(t)$ to increase linearly and exceed $E[T_N] \approx N \ln N$, in both cases approximately.
This formula gives the predicted value of the diameter within one from either the correct or probabilistically predicted value except 
for the smallest case (the 3$\times$3$\times$3 Cube in the square-turn metric), where the error is three.

In the following article \cite{Hirata2024_geom}, a lower bound for the diameter of symmetric Cayley graphs of some Rubik's Cube groups 
is obtained on the basis 
of graph-theoretical arguments, and compared with lower and upper bounds for the diameter of random regular graphs of Bollob\'{a}s and de la Vega \cite{BollobasVega1982,Bollobasbook},
who combined graph-theoretical and probabilistic arguments. 

\acknowledgments
The author is a Guggenheim Fellow of the John Simon Guggenheim Memorial Foundation. 

\appendix

\section{Coupon collector's problem \cite{Gumbel1941,Dawkins1991,Flajolet1992,Blom1994}\label{sec:coupon}}

Let us consider a process of uniformly randomly generating an integer in the range of 1 through $N$.
How many times ($T_N$) does the random generation need to be repeated before every integer in the range of 1 through $N$ occurs at least once?
Let  $E[T_N]$ be the expectation value of this number.

First, suppose that $n$ distinct integers have appeared so far after some number of random generations. 
How many additional times ($t_n$) does one need to carry out the random generations before a new, distinct integer  is added? 
Let $E[t_n]$ be the expectation value of this number.

The probability that the first randomly generated integer is distinct is
\begin{eqnarray}
p_n = \frac{N-n}{N}.
\end{eqnarray}
The probability that the first randomly generated integer is not distinct, but the second generated integer is is
\begin{eqnarray}
(1-p_n) p_n.
\end{eqnarray}
Generally, the probability that the $k$th random generation produces a new, distinct integer for the first time is
\begin{eqnarray}
(1-p_n)^{k-1} p_n.
\end{eqnarray}
Writing $q_n = 1-p_n$, the expectation value  $E[t_n]$ is then an infinite sum,
\begin{eqnarray}
E[t_n] &=& 1 \cdot p_n + 2 \cdot  q_n p_n + 3 \cdot  q_n^2 p_n + \dots \nonumber\\
%&=& \left( 1 + 2 q + 3 q^2 + \dots\right)p \nonumber\\
&=& p_n\frac{d}{dq_n} \left( q_n + q_n^2 + q_n^3 + \dots \right)  \nonumber\\
&=& p_n\frac{d}{dq_n} \frac{q_n}{1-q_n}  \nonumber\\
&=& p_n\frac{1}{(1-q_n)^2} \nonumber\\
&=& \frac{1}{p_n} = \frac{N}{N-n}.
\end{eqnarray}
The corresponding variance $V[t_n]$ is 
\begin{eqnarray}
V[t_n] &=& \Big(E[t_n^2]\Big) - \Big(E[t_n]\Big)^2 \nonumber \\
&=& \left( 1^2 \cdot p_n + 2^2 \cdot q_n p_n + 3^2 \cdot q_n^2 p_n + \dots \right) - \frac{1}{p_n^2} \nonumber \\
%&=& \left( 1 + 4 q + 9 q^2 + \dots\right)p - \frac{1}{p^2} \nonumber\\
&=& p_n\frac{d}{dq_n} \left( q_n + 2 q_n^2 + 3 q_n^3 + \dots \right)  - \frac{1}{p_n^2}\nonumber\\
&=& p_n\frac{d}{dq_n} q_n \frac{d}{dq_n} \left( q_n + q_n^2 + q_n^3 + \dots \right)  - \frac{1}{p_n^2} \nonumber\\
&=& p_n\frac{d}{dq_n} q_n \frac{d}{dq_n} \frac{q_n}{1-q_n}  - \frac{1}{p_n^2} \nonumber\\
&=& p_n\frac{d}{dq_n} \frac{q_n}{(1-q_n)^2}  - \frac{1}{p_n^2} \nonumber\\
&=& \frac{1}{p_n^2} - \frac{1}{p_n} = \frac{N^2}{(N-n)^2} - E[t_n].
\end{eqnarray}

Then, $E[T_N]$ is the sum of the average number $E[t_n]$ of random generation needed to produce the ($n$+1)th distinct integer over $0 \leq n \leq N-1$. Therefore, in the large-$N$ limit,
\begin{eqnarray}
E[T_N] &=& E[t_0] + E[t_1]+ E[t_2] + \dots + E[t_{N-1}] \nonumber\\
&=& \frac{N}{N} + \frac{N}{N-1} + \frac{N}{N-2} + \dots + \frac{N}{1} \nonumber\\
&=& N \left( \frac{1}{1} + \frac{1}{2} + \frac{1}{3} + \dots + \frac{1}{N} \right) \nonumber\\
&\approx& N \ln N + \gamma N,
\end{eqnarray}
where $\gamma$ is the Euler--Mascheroni constant. 
The corresponding variance $V[T_N]$ is
\begin{eqnarray}
V[T_N] &=& V[t_0] + V[t_1]+ V[t_2] + \dots + V[t_{N-1}] \nonumber\\
&=& \frac{N^2}{N^2} + \frac{N^2}{(N-1)^2} + \frac{N^2}{(N-2)^2} + \dots + \frac{N^2}{1^2} - E[T_N] \nonumber\\
&=& N^2 \left( \frac{1}{N^2} + \frac{1}{(N-1)^2} + \dots + \frac{1}{2^2} + \frac{1}{1^2} \right) - E[T_N] \nonumber\\
&\approx& \frac{\pi^2N^2 }{6} -N \ln N - \gamma N, 
\end{eqnarray}
where we used 
\begin{eqnarray}
\sum_{k=1}^{\infty} \frac{1}{k^2} =\frac{\pi^2}{6}.
\end{eqnarray}
For large $N$, the standard deviation $\sigma[T_N]$ is
\begin{eqnarray}
\sigma[T_N] = \sqrt{V[T_N]} \approx 1.28\,N. %\frac{\pi}{\sqrt{6}}\,N  = 1.28 \,N .
\end{eqnarray}

\section{Derivation of Eq.\ (\ref{deltaSlargen}) \label{sec:derivation2}}

Consider a process of uniformly randomly generating $n$ integers in the range of 1 through $N$.
How many ($N_n$) distinct integers are expected to be present in this set of $n$ integers? Let $E[N_n]$ be this expectation value.

Let $p_k$ be the probability that the $k$th randomly generated integer is distinct from all previously generated ones. 
Let $N_k$ be the number of distinct integers after the random generation is repeated $k$ times. 
We have
\begin{eqnarray}
p_{k+1} = \frac{N - N_{k}}{N},
\end{eqnarray}
which can be rearranged to 
\begin{eqnarray}
N_{k} = N(1- p_{k+1}). \label{nk}
\end{eqnarray}
%p_1 &=& 1, \nonumber\\
%p_2 &=& p_1 \frac{N-1}{N}  = p_1 \left( 1- \frac{1}{N} \right), \nonumber\\
%p_3 &=& p_2p_1 \frac{N-2}{N} + (1-p_2)p_1 \frac{N-1}{N} = p_2 \left( 1- \frac{1}{N} \right), \nonumber\\
%p_4 &=& p_3p_2p_1 \frac{N-3}{N} + (1-p_3) p_2p_1 \frac{N-2}{N} \nonumber\\
%&& + p_3 (1-p_2) p_1 \frac{N-2}{N} + (1-p_3)(1-p_2)p_1 \frac{N-1}{N} \nonumber\\
%&=& p_3 \left( 1- \frac{1}{N} \right).
On the other hand, 
\begin{eqnarray}
N_{k} = p_{k} \left( N_{k-1} +1 \right) + \left(1-p_{k}\right) N_{k-1} = N_{k-1} + p_k. \label{nk2}
\end{eqnarray}
Substituting Eq.\ (\ref{nk}) into Eq.\ (\ref{nk2}), we obtain
\begin{eqnarray}
p_{k+1} &=&  \left( 1- \frac{1}{N} \right) p_k  = \left( 1- \frac{1}{N} \right)^k, 
\end{eqnarray}
where we used $p_1 = 1$ and $N_0 = 0$. 
Therefore, after $n$ random integers are generated, the expected number of distinct integers among them is
\begin{eqnarray}
E[N_n]&=& N (1- p_{n+1}) \nonumber\\
%&&= \left\{ \left(1-\frac{1}{N}\right)^0 +  \left(1-\frac{1}{N}\right)^1 + \dots +  \left(1-\frac{1}{N}\right)^{n-1}\right\} \nonumber\\
&=&N  \left\{ 1 - \left(1-\frac{1}{N}\right)^n \right\} \nonumber\\
&\approx& N \left(1-e^{-n/{N}}\right),
\end{eqnarray}
in the large-$N$ limit.

For $n \ll N$, approximating the exponential by its first-order Taylor expansion, we have
\begin{eqnarray}
E[N_n] = N \left(1-e^{-n/{N}}\right) \approx N \left\{ 1 - \left(1- \frac{n}{N} \right) \right\} = n,
\end{eqnarray}
indicating that virtually all of $n$ randomly generated integers are distinct. 

The expected number of integers that are not generated even once after $n$ random generations is 
\begin{eqnarray}
N - E[N_n] = N - N \left(1-e^{-n/{N}}\right) = N e^{-n/{N}} .
\end{eqnarray}
%which is consistent with the Siobhan distribution [Eq.\ (\ref{siobhan})].
When this number is smaller than one, it is expected that all of the $N$ integers have been realized. 
\begin{eqnarray}
N e^{-n/N} < 1,
\end{eqnarray}
which implies
\begin{eqnarray}
N \ln N < n.
\end{eqnarray}
This is consistent with the solution of the coupon collector's problem given in Appendix \ref{sec:coupon}.

\bibliography{rubik.bib}

\end{document}